\documentclass[12pt]{article}
\usepackage[dvips]{graphicx,color}
\usepackage{overcite,dcolumn,amsmath,amssymb}
\newcommand{\mc}[1]{\multicolumn{1}{c}{#1}}

\definecolor{rulecolor}{gray}{0.5}
\newsavebox{\HttpBox}
\begin{lrbox}{\HttpBox}
\footnotesize\verb+http://www.mpip-mainz.mpg.de/~yasp/ccteam/index-gruppe.html+
\end{lrbox}

\begin{document}
\title{Deriving effective mesoscale potentials
       from atomistic simulations}
\author{Dirk Reith, Mathias P\"utz, and Florian M\"uller-Plathe\\
  \small Max-Planck-Institut f\"ur Polymerforschung, D-55128 Mainz,
  Germany}
\date{Draft report, \today,  for submission to JCC}
\maketitle
%
%
\begin{abstract}
  We demonstrate how an iterative method for potential inversion from
  distribution functions developed for simple liquid systems can be
  generalized to polymer systems.  It uses the differences in the potentials
  of mean force between the distribution functions generated from a guessed
  potential and the true (simulated) distribution functions to improve the effective
  potential successively.  The optimization algorithm is very powerful:
  convergence is reached for every trial function in few iterations.
  As an extensive test case we
  coarse-grained an atomistic all-atom model of poly (isoprene) (PI) using a
  13:1 reduction of the degrees of freedom. This procedure was
  performed for PI solutions as well as for a PI melt.  Comparisons of the
  obtained force fields are drawn. They prove that it is not possible to use
  a single force field for different concentration regimes.
\end{abstract}

%
\section{Introduction}
%
In polymer theory, there is still a strong focus on developing a qualitative
understanding of many fundamental problems like polymer dynamics in different
environments, crystallization or the thermodynamics of complex polymer
systems. Most computational studies of specific polymers are based on
atomistic models and tools developed to address specific local aspects of
polymer behavior. While these methods are quite useful to gain qualitative
insight, quantitative predictions of the behavior of entire chains are very
hard to obtain from them.  \\ Although computational power uses to increase
ten-fold every 5 years~\cite{vangunsteren:90}, the huge number of degrees of
freedom limits brute force approaches when it comes to investigating phenomena
at meso- and macroscopic length scales.  One way to circumvent this problem is
to reduce the degrees of freedom by coarsening the models and keeping only
those degrees of freedom which are deemed relevant for the particular range of
interest. Simple models for the study of meso- and macroscale phenomena in
polymers have been used extensively~\cite{degennes:79,doi:86}. Because of
their generic nature, however, most of them do not distinguish between
chemically different polymers. The present contribution describes a method for
systematically generating mesoscale models from atomistic models. The
mesoscale or coarse-grained (CG) models contain enough information to retain
the chemical identity of the parent polymer.  Thus, they are able to describe
specific polymeric systems.  \\ The idea of coarse graining is not
new~\cite{baschnagel00,bai_new,fmp_new}.  To highlight a few approaches, Murat
and Kremer \cite{murat98} mapped bead-spring type polymer chains in a melt to
a soft-core liquid with fluctuating ellipsoidal particles modeled by an
anisotropic Gaussian potential. A similar way to coarsen polymer chains in
dilute to semi-dilute solutions was recently found by Louis et
al.\cite{bolhuis00s,louis01s}. Closest to our approach, Tsch\"op and
successors \cite{tschoep98a,hahn01} and Akkermans et al.\cite{akkermans01} did
systematic studies of polymer melt coarse graining. However, none of these
derived an automatic optimization scheme to conserve the chemical nature of
the underlying atomistic model in a standardized way as we do.  \\
Because of the similarity of the coarsened polymer model to simple liquids one
might ask if the well-developed arsenal for simple liquids can solve the
inversion problem of going from measured quantities like pair correlation
functions to effective model potentials. The polymer connectivity and the
fact, that we deal with coarsened "super-atoms" complicates the matter
significantly. Nonetheless, the self-consistent polymer reference interaction
site model (PRISM) theory, developed by Schweizer and Curry
\cite{schweizercurro89}, combined with Monte-Carlo methods was shown to be
able to qualitatively predict both short- and long-range structure of polymer
chains in melts \cite{weinholdcurro99,puetz01}. So far, a route to apply PRISM
theory to obtain coarse-grained potentials has not been established.  Reverse
Monte-Carlo (RMC) techniques \cite{rmc,mcgreevy95,Sop96cp} simulate a particle
system to produce the correct radial distribution function without the
explicit need for a potential. The lack of a potential though limits the
usefulness of the approach.
As explained in reference \cite{revans90}, RMC methods are useful
for obtaining higher order correlation functions from the
knowledge of the pair distribution functions, without knowing the
underlying forces. Attempts similar in spirit to RMC have also
been developed by Lyubartsev and Laaksonen~\cite{lyubartsev95} and
later been applied to DNA studies~\cite{lyubartsev99}.
\\ From a practical point of view, it is desirable to establish a standard
framework for coarse-graining of polymer systems, i.e. a formalism
which allows one to calculate a broad range of properties with a
minimum amount of manual interference.  We develop an automatic and
iterative way to determine effective interactions which properly match
a set of quantities calculated from a higher detailed reference
simulation model (i.e.\ atomistic).  While in principle one could also
use experimental reference data (as has been done in the related
problem of atomistic force field optimization~\cite{faller99}), the
structural distributions needed here are difficult or impossible to
measure with sufficient accuracy.  \\To map atomistic to mesoscopic
models, one groups several atoms together into "super-atoms". On
this coarsened length scale, effective potentials are sought in an
optimization procedure which reproduce the structural
distributions between the super-atoms, as obtained from a simulation of
the atomistic model. Reith et al.\cite{reith00s} successfully applied
a simplex algorithm to such an optimization. However, slow
convergence of the analytical potentials and the manual process of
selecting a good functional form of the potential are drawbacks of
this method. The new approach presented here removes some of the
ambiguity by using tabulated numerical potentials instead. The method
is called iterative Boltzmann inversion. \\ It shall be introduced 
as follows: In section \ref{sec:method} we derive our
inversion method for polymer models.  Then (section
\ref{sec:convergence}) we test the convergence of the inversion scheme
on a simple Lennard-Jones liquid and discuss ways of improving the
convergence rate.
In section~\ref{PI} we show how the method is applied to
obtain CG potentials from atomistic simulation data for the example of poly
(isoprene). The resulting CG model reproduces a chosen set of structural
details of the atomistic system to very high accuracy.
%
%
\section{Iterative Boltzmann Inversion}
\label{sec:method}
\begin{figure}
  \begin{center}
    \includegraphics[width=0.56\linewidth]{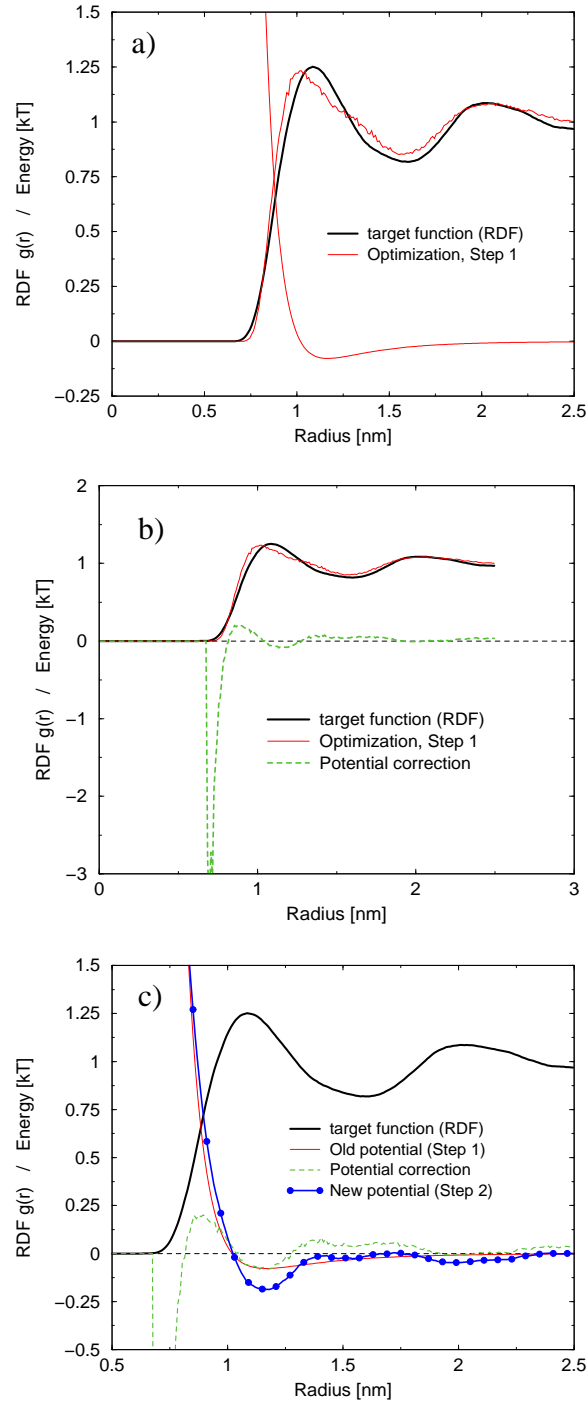}
  \end{center}
  \caption{Illustration of the iterative procedure. (a): A test simulation (step 1)
    is executed with an arbitrary chosen initial potential. This results in a
    trial RDF for this first optimization step. (b): Taking the ratio between
    trial and target RDF leads via Boltzmann-inversion to a potential
    correction. (c): By adding the correction to the old potential, the new
    potential for the next iteration (step 2) is generated.  }
  \label{fig:opt_scheme}
\end{figure}

%
The general problem of creating a realistic model consists of
writing down a Hamiltonian which can then be parametrized to
reproduce empirical information about a system as well as
possible.  To define the Hamiltonian, one needs to choose a set of
particles and interactions between them. The choice of these variables
(in our context CG super-atoms, how many of them, their relation to 
the underlying real atoms) can hardly be automated and a good choice will
probably always depend on ones physical intuition. Hence, although
important, we pay little attention to this part of the problem in this work.
The interactions between the particles have most commonly
been chosen to be analytic functions with a few adjustable
parameters. Observables can usually
be expressed as functionals of correlation functions
between the various degrees of freedom.  It has been argued that
if all interactions with potentials $V^{(n)}(r_1,...,r_N)$ ($x_i$
labeling the degrees of freedom) in an atomistic system consist of
$n$-body and lower terms, then the system is completely determined
by the knowledge of all correlation functions
$g^{(n)}(x_1,...,x_N)$ of $n$-th order and lower
(cf.\ Reference~\cite{zwicker90}). In practice a complete determination of
$n$-point ($n > 4$) correlation functions for $N > 2$ particles is
a huge task for all but the simplest model cases and one typically
has to restrict oneself to a limited subset. Even if one knew all
relevant correlation functions for a statistical mechanical
system, one would still have to face a high-dimensional
non-linear inversion problem to derive the set of effective
potentials.  Here, we limit ourselves to correlation
functions which only depend on a single coordinate like radial
distribution functions ($g(r)$), bond distance
($d(r)$), bending angle ($a(\alpha)$), and dihedral
angle distributions ($b(\beta)$).  These distribution functions
and variables are a convenient choice to describe the structure of
polymers, since they
allow a separation of intra- and intermolecular structure which is
naturally motivated by the chemical connectivity of polymer
chains. They enable the use of an iterative
Boltzmann inversion scheme to extract effective potentials from a
set of known correlation functions.
\\
Let us illustrate the procedure with the example of deriving an effective
non-bonded potential $V_0(r)$ from a given radial distribution function $g(r)$.  
We first need a reasonable initial guess.  It has been proposed~\cite{Sop96cp} to
invert radial distribution functions for one-component simple liquid systems
by taking a simple Boltzmann inverse of $g(r)$. This is, however, exact only
in the limit of infinitely dilute systems, i.e. with density $\rho=0$. 
We use the potential of mean force 
\begin{equation} \label{eq:boltzinv}
  F(r) = -k_B T \ln g(r) \, ,
\end{equation}
which is a free energy and not a potential energy (except for the uninteresting
case of zero density). However, $F(r)$ is usually sufficient to serve as the 
initial guess $V_0(r)$
for an iterative procedure. (Other choices are possible as well, e.g.\ hard
spheres.) Simulating our system with $V_0(r)$ will yield a
corresponding $g_0(r)$ which is different from $g(r)$. The potential
needs to be improved, which can be done by a correction term $- k_B T
\ln \left( \frac{g_0(r)}{g(r)}\right)$. This step can be iterated:
\begin{equation}
 V_{i+1}(r) = V_i(r) - k_B T \ln \left( \frac{g_i(r)}{g(r)}\right) \, .
\end{equation}
The procedure is illustrated in Figure~\ref{fig:opt_scheme}.
Clearly a potential that reproduces $g(r)$ is a fixed point of the
iteration. Thus, if the algorithm converges we have a valid solution.
Soper \cite{Sop96cp} gives some qualitative arguments why one can
expect convergence. In general, each iteration tends to
over-correct the current potential as we demonstrate for simple
Lennard-Jones test case in the next section.
\\ A generalization to more than one potential/distribution function pair is
straightforward. We simply have to replace the potential $V(r)$
and distribution function $g(r)$ by the appropriate pair of
functions, e.g. $V_{bend}(\theta)$ and $a(\theta)$ in case of an
intramolecular bending interaction. Since in dense systems
individual distributions usually depend on the full set of
potentials through higher-order correlations, one cannot simply
iterate for each potential separately. Although one can keep all
other potentials constant while iterating a particular one, one
must re-adjust after other potentials are changed. For practical
purposes, one should start with those potentials which are least
affected by changes to all other ones, e.g. bending potentials
before non-bonded potentials.  The speed of convergence can be
influenced strongly by the order in which one optimizes the
various potentials and, in the case of non-bonded interactions, by
limiting the range of the potentials. The latter will be
demonstrated in depth in the following section. All details of
implementation have been published separately,~\cite{reith_crc02}
in order to streamline the flow of the physical arguments.

%
%
\section{Convergence Tests with Model Liquids}
\label{sec:convergence}
%
For the method to be useful, it should converge rapidly, the
resulting potential should be physically sensible, and it should
find the correct potentials in known test cases. An additional
consideration arises from the fact that the same radial
distribution function can be reproduced to within its uncertainties 
by several visibly different potentials, as will be shown below.
Hence, it is useful to establish criteria for selecting a specific
one from a set of potentials.\\

Before trying polymers, we therefore tested the iteration scheme
with two dense model liquids ($\rho^* = 0.85$, $T^* = 1$ in
reduced units~\cite{All87}): we applied firstly, the purely
repulsive Weeks-Chandler-Andersen  (WCA) potential~\cite{All87} 
which equals zero for $r > 2^{1/6}\sigma$ and secondly, the
Lennard-Jones(6-12)-(LJ)potential, with a cutoff distance of
$r_{\rm cut} = 2.5 \sigma$. Both systems contained $N=1024$
particles and were simulated in the $NVT$ ensemble.
\begin{figure}
  \begin{center}
    \includegraphics[width=0.7\linewidth]{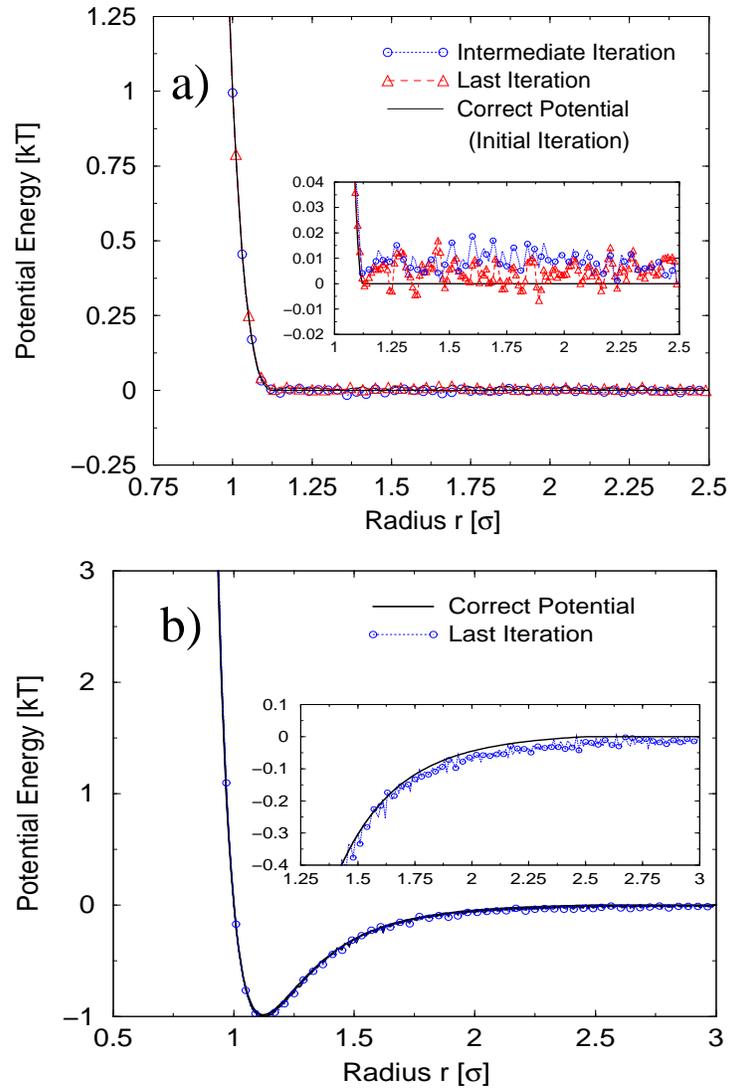}
  \end{center}
  \caption{Stability test of iteratively optimized potential functions
  for a dense liquid of (a) WCA and (b) Lennard-Jones model atoms. Although
  being short-ranged ((a): $\approx 1.1\sigma$, (b): $2.5\sigma$), the
  true potentials were chosen as initial guesses of a long-ranged artificial
  optimization. In each step, a five-point running average was applied to both
  the trial RDF and to the new potential. The potential shape
  remains the same during a total of $35$ iteration steps; for reasons of
  clarity, only a few iterations are shown. It should be noted that all 
  potentials shown here reproduced the corresponding radial distribution
  functions to within line thickness.}
  \label{statio}
\end{figure}
As target function to be reproduced, we chose the radial
distribution functions (RDF) corresponding to the non-bonded WCA
or LJ potential, respectively. We want to emphasize the
point, that here we were in the comfortable but rare situation to
know \textsl{exactly} the correct potential function(s) and, hence,
the solution(s) of the inversion problem. Each
system was simulated for $10^5\tau$ ($\tau$ is the
LJ-time~\cite{All87}) to obtain smooth RDFs, which we calculated
in the range $r \in [0; 4.5\sigma]$ with $\Delta r = 0.01\sigma$.
\\

Firstly, we tested the numerical stability of our iterative
potential solver. In both cases, we used the correct 
potential with an extended range as input, padding it with zeros.
In case of the LJ liquid, the extension reached from $2.5\sigma$
to $3\sigma$, in case of the WCA liquid from $\sqrt[6]{2}\sigma$ 
to $2.5\sigma$. I.e. we tested how far the algorithm moves away if
started with the perfect solution. We observed that
the iterations introduced non-zero values (below $\pm 0.02 k_BT$)
of the potential beyond the range of the true potential which are
clearly artifacts. They could be nearly removed by smoothing both
the trial RDFs and the iteratively changed trial potentials.
Without smoothing, the statistical noise is too large to
conserve the initial potential shape for $r > r_{\mbox cut}$. But
a five-point running average was sufficient to maintain the correct 
shape as shown in Figure~\ref{statio}. Part (a) shows the results
for the WCA potential.
Even after $35$ iterations, the energy fluctuations at large 
$r > r_\text{cut}$ are below $0.005 k_BT$, being in the 
same order of
magnitude as the fluctuations from an intermediate run. Similar 
deviations were found for the LJ liquid (Figure~\ref{statio}b). The
problem is that unphysical fluctuations on the scale of the mesh
are not sufficiently suppressed for large $r$ within the
accuracy of our calculations. A strong positive peak in the potential
at a particular mesh point countered by a negative peak in directly neighboring
mesh point disturbs the resulting RDF only very weakly, especially if they occur
at large $r$. Hence they have a tendency to destabilize the procedure
and therefore have to be suppressed by additional means, i.e. some
convenient smoothing algorithm. This could be a problem if the RDF structure
shows strong gradients. Then the smoothing algorithm has to be chosen
very carefully to conserve the real physical features.
\begin{figure}
  \begin{center}
    \includegraphics[width=0.56\linewidth]{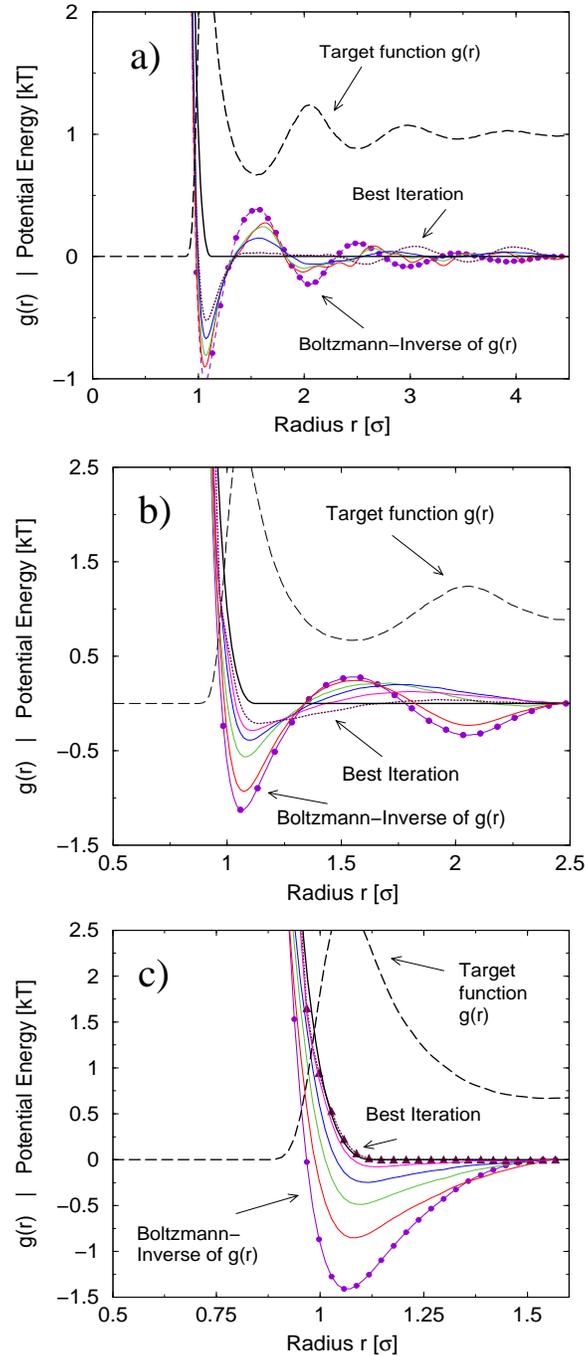}
  \end{center}
  \caption{Potential reproduction test for a dense liquid of WCA model
  particles. In all cases, the iteration starts from the Boltzmann-
  inverted potential of the target RDF. The range of the numerical potential 
  is (a) $4.5\sigma$, (b) $2.5\sigma$, (c) $1.6\sigma$. For reasons of
  clarity, graphs from only some iteration steps are drawn. In every case,
  the target RDF can be quickly reproduced, but only for (c) the converged
  potential is purely repulsive.}
  \label{reprod}
\end{figure}
\\ 
Secondly, we investigated if the correct potential is retrieved when
we start the iterations from the Boltzmann-inverted function of
the target RDF. These optimizations are done for various ranges of
$r$ in order to check what happens if the range does not match the
range of the true potential. Starting from the Boltzmann-inverted
potential, the target RDF is not matched immediately, but 
$3 - 10$ iterations yielded the correct shape (RDFs match within
their fluctuations). This underlines the robustness of the method.

The situation is not quite as favorable for the corresponding potential
functions. For the WCA system, the results are given in 
Figure~\ref{reprod}. In all cases,
the optimization starts from the Boltzmann-inverted potential of
the target RDF. The $r$-ranges for the numerical potential were chosen
according to the position of the minima in the target RDF: (a)
$4.5\sigma$, (b) $2.5\sigma$, and (c) $1.6\sigma$. We picked the
minima because in the valleys we expect fewer particles which have 
to experience the force discontinuity when
entering the non-zero range of the potential. For case (a) we executed $15$
iterations until we decided, that the long-ranged fluctuations we
unlikely to disappear in an acceptable number of iterations, even 
though there was
a tendency to reduce the depth of the first potential minimum
($0.5 k_BT$). Case (b) was iterated for $28$ steps and the result
was much better: the depth of the first minimum could be
significantly reduced ($0.25 k_BT$) and for $r > 1.5\sigma$ the
potential decayed almost to zero. However, further progress was
slow so that we stopped the optimization.  In the last case (c),
we could reach the original potential shape within line thickness
after $13$ steps and it remained stable thereafter, as expected
from the previous test.\\
\begin{figure}
  \begin{center}
    \includegraphics[width=0.7\linewidth]{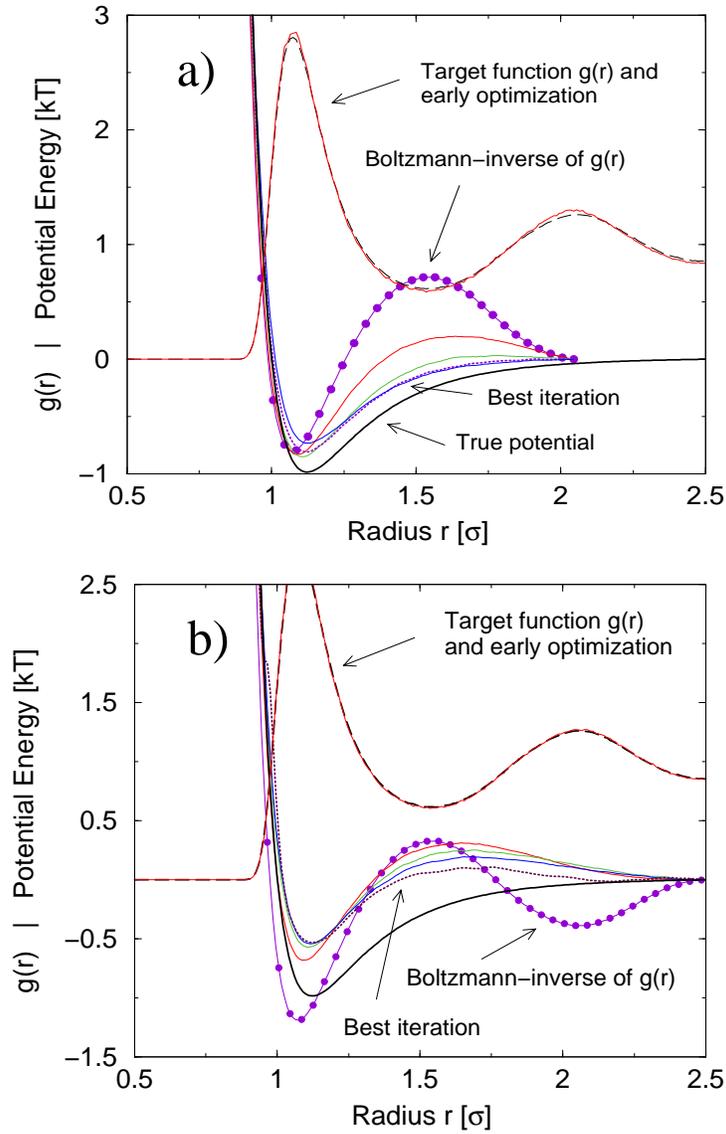}
  \end{center}
  \caption{Potential reproduction test for a dense liquid of LJ model
  particles. In both cases, the optimization starts from the Boltzmann-
  inverted potential of the target RDF. The range of optimization is
  (a) $2.05\sigma$, (b) $2.5\sigma$. For reasons of
  clarity, graphs from only some iteration steps are drawn. In every case,
  the target RDF can be quickly reproduced, however, the best iterative
  potentials do not exactly match the original function. By watching the
  slope note, that the derivatives (i.e. the forces) match very well up to
  $r \approx 1.5\sigma$.}
  \label{eindeut}
\end{figure}
For the Lennard-Jones system, we tried two different ranges of
optimizations (cf. Figure~\ref{eindeut}): (a) $2.05\sigma$, 
(b) $2.5\sigma$. In case (a) we
could see how robust the method was with an unfavorable cut-off:
first we chose a shorter range than the true one and then we chose to
cut it of at a peak value in the RDF, generating (with the initial
Boltzmann-inverted RDF as the potential guess) the situation
that particles coming into the range of the potential first
experience a repulsion instead of the physically meaningful London
attraction.  The result was as expected: the artificial repulsive
part around $1.5\sigma$ had vanished completely and the slope of 
the final potential
almost matched the slope of the true potential, i.e. the forces
are very similar. In fact, this result was even better than what
we obtained for the correct range in case (b). Here, the
attraction at around $r \approx 2\sigma$ slowed down the
convergence and the decrease of the non-physical repulsion at
around $r \approx 1.5\sigma$. Up to this distance, however, the
forces were nicely matched, too. Taken together, our new iteration
scheme satisfies the demands of stability and qualitative
reproduction of the true potential function. However, in practice
there can be many (similar) potential functions which can
reproduce a given target RDF, even in the simple case of
mono-atomic dense liquid systems. For practical as well as for
physical reasons, it makes sense to choose the shortest-range
potential, that reproduces the RDF, from a selection of
possibilities. Alterations may be done thereafter by increasing
the potential and RDF range successively in a series of
optimizations until one is satisfied with the reproduction of the
target RDF. Note also, that we did not yet explore the field of
well-known convergence accelerators (e.g. shifting cutoff or
dynamic prefactors for the correction term~\cite{Sop96cp}) except
for our simple working solutions.

%
\section{Poly (isoprene) - Coarse Graining of a realistic model}
\label{PI}
%
All coarse-grained force fields depend on system properties like density,
temperature, and composition. Therefore, for two different concentrations, 
a melt and dilute solution, two force fields had to be
constructed. The choice of the mapping centers was identical.
Therefore, we first describe the technical details and the mapping
procedure common to both systems. Subsequently, the specific
optimizations will be discussed.
\subsection{Technical Simulation Details}
The atomistic simulations which are our starting point for the coarse graining
are described in detail in Reference~\cite{faller01s} in case of the melt and
in Reference~\cite{fallerreith02} in case of the solution. We only briefly
summarize the main characteristics. The melt simulation contains 100 oligomers
of length 10 monomers at a density of $890$~kg/m$^3$. All chains are {\it
  trans}-polyisoprene (cf.\ Figure~\ref{polyisoprene}). We applied a
self-developed all-atom force-field resulting in 132 interaction
sites or atoms per chain. The atomistic simulation lasted $1.1$~ns
at ambient conditions ($T=300$K, $p=101.3$kPa). This model 
describes the relaxation of local time correlation functions in agreement
with NMR measurements and it reproduces the melt structure factor of
poly (isoprene)~\cite{faller00_phd} reasonably well.\\

The solution simulation was done with one single strand of 15
monomers dissolved in cyclohexane. We chose ambient conditions
with a density of $763$~kg/m$^3$ and a polymer concentration of
weight $4.6$~\%. The polymer force field was the same as in the
melt case except that only interactions up to $1-4$ interactions
were excluded. Some thermodynamic and static properties of the
atomistic systems are listed in Table~\ref{at-cg-props}, following
the nomenclature applied in Reference~\cite{fallerreith02}.\\ 

For the CG simulations, both MD (for the melt) and Monte Carlo (MC,
for the solution) programs have been used. All MD runs are
performed in the \textit{NVT} ensemble. The system consists of an
orthorhombic box with periodic boundary conditions. The Langevin
equations of motion are integrated by the velocity Verlet
algorithm with a time step $\Delta t = 0.01\tau$.\cite{All87} A
mean temperature of $k_{B}T=1$ is maintained by the Langevin
thermostat with friction constant $\Gamma = 0.5
\tau^{-1}$.\cite{grest86} For the MC simulations, the single chain
program \textsl{prism}~\cite{puetz01} is applied. It applies
various kinds of Pivot moves to simulate a canonical ensemble. We
carry out $10^5$ attempted warm-up moves before a production run of
$10^6$ attempted MC-moves is started (acceptance rate varied typically
between 5 and 10\%).
\begin{table}
  \begin{center}
    \caption{Characteristic properties of the atomistic poly (isoprene)
      (PI) simulations. The systems labels are following the nomenclature applied
      in Reference~\cite{fallerreith02}. $N_P$ is the number of PI oligomers
      (10-mers in the melt and 15-mers in solution) and
      $N_C$ the number of cyclohexane molecules. $c$ is the concentration in
      weight \% polymer. $t_{\text{sim}}$ is the simulated time for the
      systems. $R_H$ refers to the hydrodynamic radius, $R_G$ to the radius of
      gyration, and $R_e$ to the end-to-end distance of the PI oligomer strands.}
    \begin{tabular}{ccccccccc}
      \hline
      system & $N_P$ & $N_C$ & c & \mc{$\rho$}\text{[kg/m}$^3$] &
      $t_{\text{sim}}$ [ns] & $R_H$ [nm] & $R_G$ [nm] & $R_e^2/R_G^2$ \\
      \hline
      M1 & 100 & 0 & 100\% & 890.0 & 1.18 & 1.13 $\pm$ 0.07 & 0.76 $\pm$ 0.05
      & 6.0 \\
      S1 & 1 & 250 & 4.6\% & 764.2 & 11.25 & 1.33 $\pm$ 0.12 & 1.21 $\pm$ 0.20
      & 6.1 \\
      S3 & 2 & 500 & 4.6\% & 762.5 &  7.81 & 1.34 $\pm$ 0.10 & 1.23 $\pm$ 0.20
      & 6.5 \\
      \hline
    \end{tabular}
    \label{at-cg-props}
  \end{center}
\end{table}
\subsection{The Mapping Procedure}
%
%
\begin{figure}[htbp]
  \begin{center}
    \includegraphics[width=0.7\linewidth]{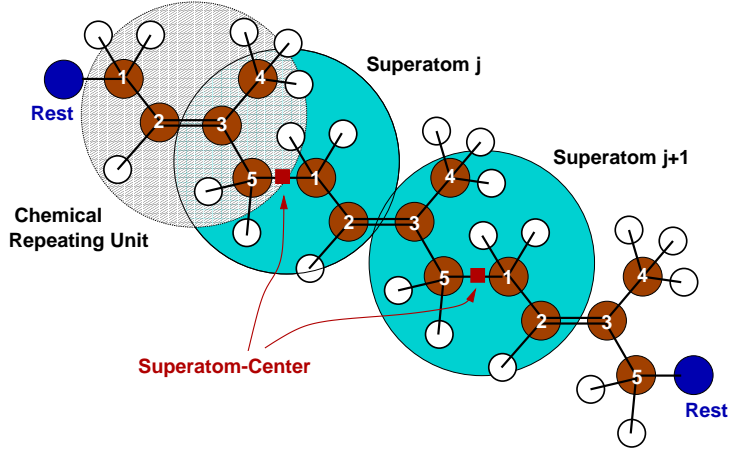}
  \end{center}
  \caption {Illustration of the mapping of
    \textit{trans}-1,4-poly (isoprene) from the atomistic to the mesoscopic
    level. Each chemical repeat unit is represented by one super-atom. As
    center of these super-atoms, we choose the middle of the atomistic bond
    between two successive chemical repeat units. This is useful because the
    resulting mapped chains generate well-defined peaks for the
    various intramolecular distributions.}
  \label{polyisoprene}
\end{figure}
\begin{figure}[htbp]
  \begin{center}
    \includegraphics[width=0.7\linewidth]{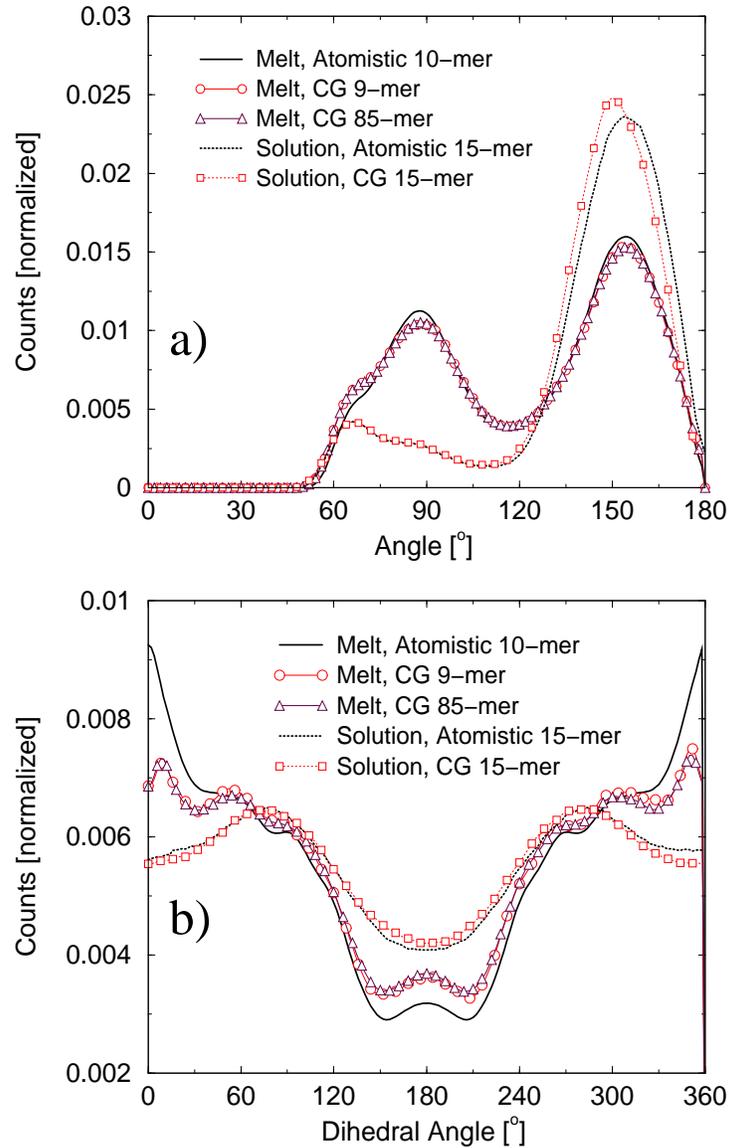}
  \end{center}
  \caption{Histogram of (a) the angle between three successive
    coarse-graining (CG) points (b) the dihedral for four successive CG points
    from atomistic and mesoscopic simulations of a poly (isoprene). Melts as
    well as solutions are considered. In both cases (a) shows that the three
    peaks from the atomistic data (centered $\approx 70^\circ$, $90^\circ$ and
    $160^\circ$) are well-reproduced by the CG simulations. In both graphs, the
    values are regularly off by not more than $10$~\%, except for some areas in
    the case of the melt torsions ($\approx 20$~\%). No dependence on the chain
    length can be observed for the melts.}
  \label{pip-intra-distr}
\end{figure}
\begin{figure}[htbp]
  \begin{center}
    \includegraphics[width=0.7\linewidth]{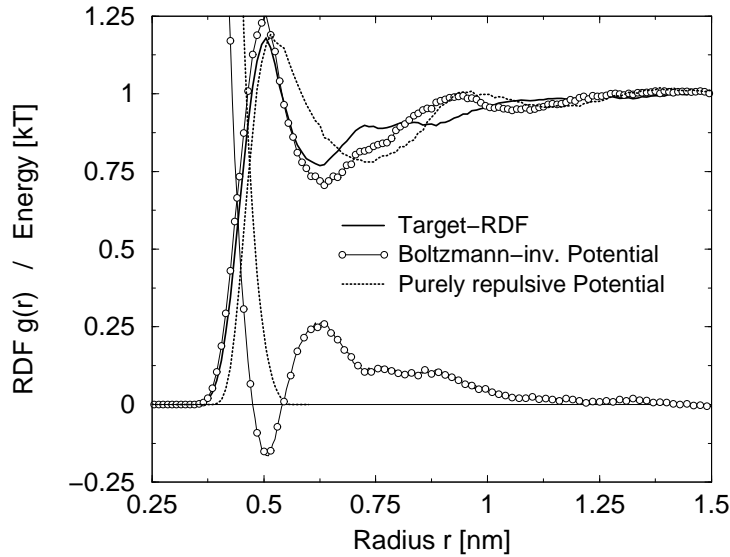}
  \end{center}
  \caption{Interchain radial distribution functions (RDFs) for super-atoms as
    obtained from atomistic simulations of a poly (isoprene) melt (9-mers) and for
    initial guesses of coarse-graining potentials: The RDFs which corresponds
    to a simulation with the potential of mean force and a purely repulsive
    potential is shown. Compared with the target RDF from the atomistic
    simulations, they can both reproduce the rise of the main peak and the
    convergence against $1$ for $r \gtrsim 1.25$~nm.  However, the
    intermediate regime is badly mimicked.}
  \label{pip-prelim-opt}
\end{figure}
\begin{figure}[tbp]
  \begin{center}
    \includegraphics[width=0.7\linewidth]{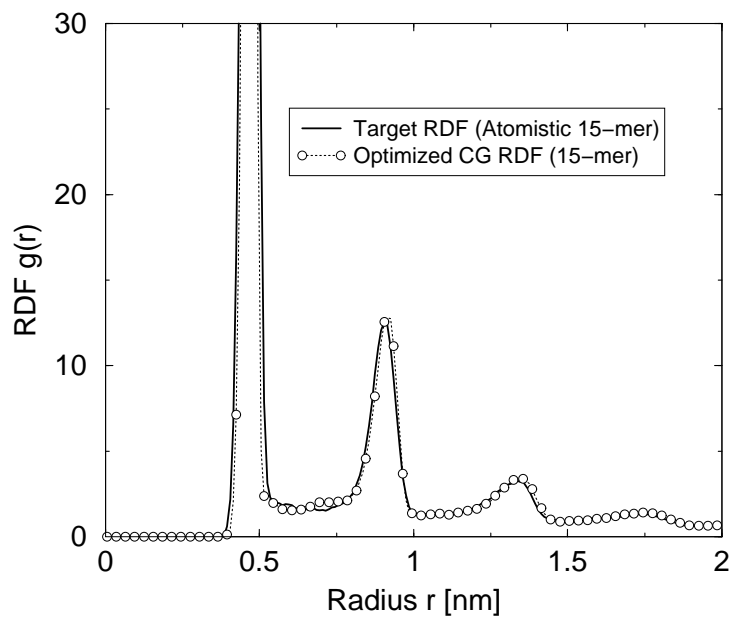}
  \end{center}
  \caption{Intrachain radial distribution functions (RDFs) for super-atoms as
    obtained from atomistic and mesoscopic simulations of a poly (isoprene)
    solution. In case of the latter, the result of a coarse-graining
    optimization is shown. The coincidence of the curves is excellent.}
  \label{rdf-opt-solv}
\end{figure}

The mapping to the mesoscale is sketched in
Figure~\ref{polyisoprene}. Each chemical repeat unit is replaced
by one super-atom, centered at the middle of the atomistic single
bond between two successive chemical repeat units. Consequently,
an atomistic $N$-mer is coarse grained to a $N-1$-mer,
disregarding the two remainders at either end. This mapping
scheme generates well distinguishable peaks
for the various intramolecular distributions. This
results from the nature of the chemical bonds of the backbone:
Every double bond is followed by three single bonds. If one chose
the center of mass of the chemical monomer, the connection between
successive super-atoms would be mediated by the C$_5$-C$_1$ single
bond. Our choice separates two super-atoms by the
C$_2$-C$_3$ double bond, which has negligible configurational
freedom, and leads to sharper peaks on the mesoscale. Moreover, the
chemical repeat unit centered around the double bond is
flat; all carbon atoms of one monomer lie in one plane. Our studies
on low-molecular weight liquids showed, that such conditions are
unfavorable for CG procedures in which the super-atoms are
represented by spheres~\cite{meyer00}. In contrast, the
C$_5$-C$_1$-centered super-atoms includes three single bonds,
which makes it more spherical. Resulting distributions for
intramolecular degrees of freedom are shown in
Figure~\ref{pip-intra-distr}, both for the melt and for the
solution. For the latter, the graphs are arithmetic means of all
three PI oligomer strands of the systems S1 and S3, for
statistical reasons (for details and nomenclature, 
compare reference~\cite{fallerreith02}).
\begin{figure}[tbp]
  \begin{center}
    \includegraphics[width=0.7\linewidth]{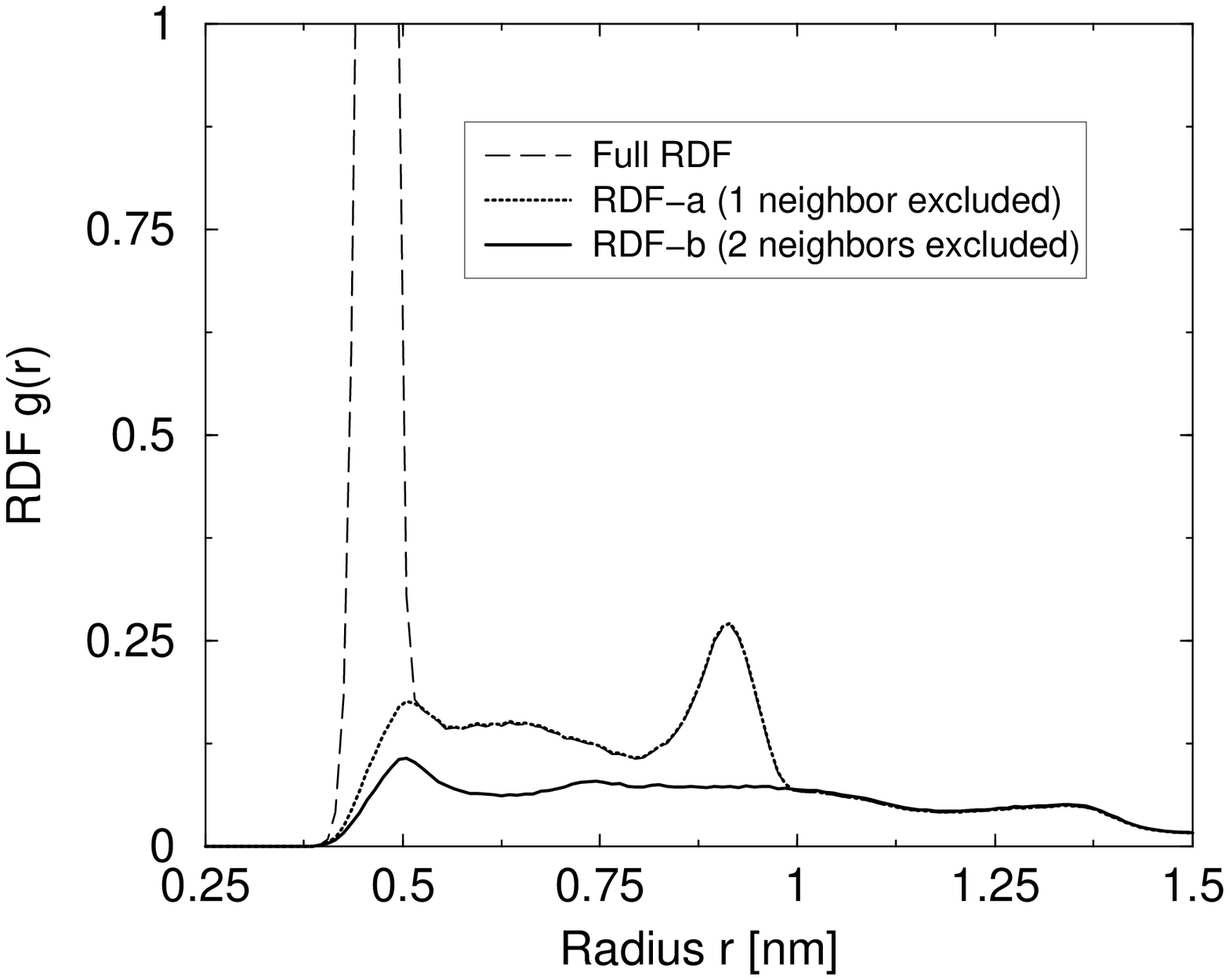}
  \end{center}
  \caption{Intrachain radial distribution functions (RDFs) for super-atoms as
    obtained from atomistic simulations of a poly (isoprene) melt (9-mers).
    RDF-a refers to an intrachain RDF for which we ignored the nearest neighbor
    chain particles. RDF-b denotes the intrachain RDF excluding the two next
    neighbors.}
  \label{at-intra-pip}
\end{figure}
Histogram (a) represents the angle between three successive
CG super-atoms , (b) the dihedral for four successive
super-atoms. They indicate that the chains are more curved in the 
melts than in solution: there is a larger proportion of angles around
$90^\circ$, in addition to the majority of dihedral angles being
around $360^\circ$, which in our notation means that the chain is
folded back onto itself ($180^\circ$ corresponds to the
fully-stretched \textit{trans} conformation).\\ The distribution of
distances between two adjacent super-atoms is well approximated by a
Gaussian, centered at $\bar{m}=0.469$~nm with a standard deviation
of $\sigma_{\bar{m}}=0.015$~nm (not shown here). For the bond stretching, 
we therefore used a simple harmonic potential. In addition to the full 
intramolecular RDF, we also
calculated the so-called RDF-a and RDF-b.  RDF-a refers to an
intrachain RDF ignoring the first neighbors along the chain. RDF-b
denotes the intrachain RDF excluding first and second neighbors.
The difference between the full RDF and RDF-a (cf.\
Figure~\ref{at-intra-pip}) shows that the main peak ($0.469$~nm)
originates from adjacent super-atoms. The difference between the
RDF-a and RDF-b reveals that the second peak around $0.9$~nm
mainly comes from the second neighbors. It corresponds to the peak
close to $160^o$ in Figure~\ref{pip-intra-distr}(a). The second
peak around $90^o$ in this figure gives rise to a hump at around $0.65$~nm 
in Figure~\ref{at-intra-pip} which it is less pronounced.  The rise of 
the RDF-b at very short distances and its substantial intensity compared 
to the RDF-a ($r \approx 0.4 - 0.8$~nm) supports the statement that the
chains are strongly curved, because such short distances are realized by 
intrachain contacts beyond second neighbor distances. The intermolecular 
RDF, as given in Figure~\ref{pip-prelim-opt} (solid line, "target RDF"), 
provides information on intermolecular contacts of the super-atoms. Major
observations are the narrowness of the main peak at $r \approx
0.5$~nm and the lack of any further strong peaks.\\ 
The situation for the solution systems is slightly different.
Figure~\ref{pip-intra-distr} indicates that the chains are much
more stretched out, especially since the fraction of bond angles
around 160$^o$ dominates greatly over all other states. This
picture is also visible in the intramolecular RDF,
given in Figure~\ref{rdf-opt-solv}. Clear peaks of first ($\approx
0.47$ nm), second ($\approx 0.9$ nm $\approx 2*0.47$ nm), and
third ($\approx 1.3$ nm $\approx 3*0.47$ nm) neighbours
corresponding to stretched-out states are separated by
low-populated areas in between. Most importantly, both the melt
and the solution system deliver nicely peaked distribution
functions with good statistical accuracy, providing a good
starting point for the coarse graining.
\subsection{Poly (isoprene) Melt Optimization}
\label{melt-opt}
As target functions for the iterative Boltzmann inversion, both the 
intra- and intermolecular RDFs were chosen. The \textit{physically} 
meaningful way to proceed has been described in full detail in a foregoing
publication~\cite{reith00s}. A viable course of action is to successively
adjust the terms contributing to the total force field in the
order of their relative strength:
\begin{equation}
  \label{eq:2}
  V_{\rm stretch} \quad\rightarrow\quad V_{\rm bend} \quad\rightarrow\quad %
  V_{\rm non-bonded} \quad\rightarrow\quad V_{\rm dihedral}
\end{equation}
%
In contrast to the technique applied in
Reference~\cite{reith00s}, the simplex algorithm, we now want to
do the optimization with the iterative Boltzmann method.
\subsubsection{Structure optimization}
The stretching potential $V_{\rm stretch}$ was considered first. 
The Gaussian shape of the distribution function turned out to be 
sufficiently mimicked by a Boltzmann-inverted potential without 
further optimization, as in the case
of poly (acrylic acid) in solution~\cite{reith00s}. So, this
degree of freedom can be treated independently from all others.
Next came the bending potential $V_{\rm bend}$. As initial guess
(as for all other intramolecular force field terms), the
Boltzmann-inverted distribution $P(\alpha)$ was applied, i.e.\
\begin{equation}
\label{v0}
    V_{\rm bend}^0(\alpha) = -k_B T \ln (P(\alpha) / \sin \alpha)
    \; .
\end{equation}
Then, we did a quick check of the distribution of the dihedral
angle and found, that the coincidence of the curves was already 
acceptable. Given the low 
importance of torsions, we therefore chose to apply already now a 
potential derived similarly to the one for the bending angle. 
\begin{equation}
\label{v1}
    V_{\rm dihedral}^0(\beta) = -k_B T \ln P(\beta) \; .
\end{equation}
Observe, that the sinus-function originating from the Jacobi-transformation
matrix when using spherical instead of cartesian coordinates, only
appears for the half-sphere bending potential. A review of the dihedral
distribution after optimizing the non-bonded part showed, that it was not 
altered significantly. The result can be viewed in 
Figure~\ref{pip-intra-distr}. Similar conclusions were drawn by 
Tsch{\"o}p et al.\ when deriving a CG force field for
polycarbonates~\cite{tschoep98a}.\\ 

Therefore, we turned to the most
crucial step in the case of melts: optimizing the intermolecular
interactions $V_{\rm non-bonded}$. To get a feeling for the correct
curvature of the intermolecular potential, we started with two
different initial choices: First, a purely repulsive potential
(WCA~\cite{weeks71}) and second, the Boltzmann-inverted potential of 
the target RDF. The results are presented in Figure~\ref{pip-prelim-opt}.
Both reproduce the rise of the main peak and
the convergence toward $1$ for $r \gtrsim 1.25$~nm. The
intermediate regime of $g(r)$ is badly reproduced. This was expected as
higher order correlations are not well captured with effective pair
potentials. We chose to start the iteration with the Boltzmann inverse.
\begin{figure}[htbp]
  \begin{center}
    \includegraphics[width=0.9\linewidth]{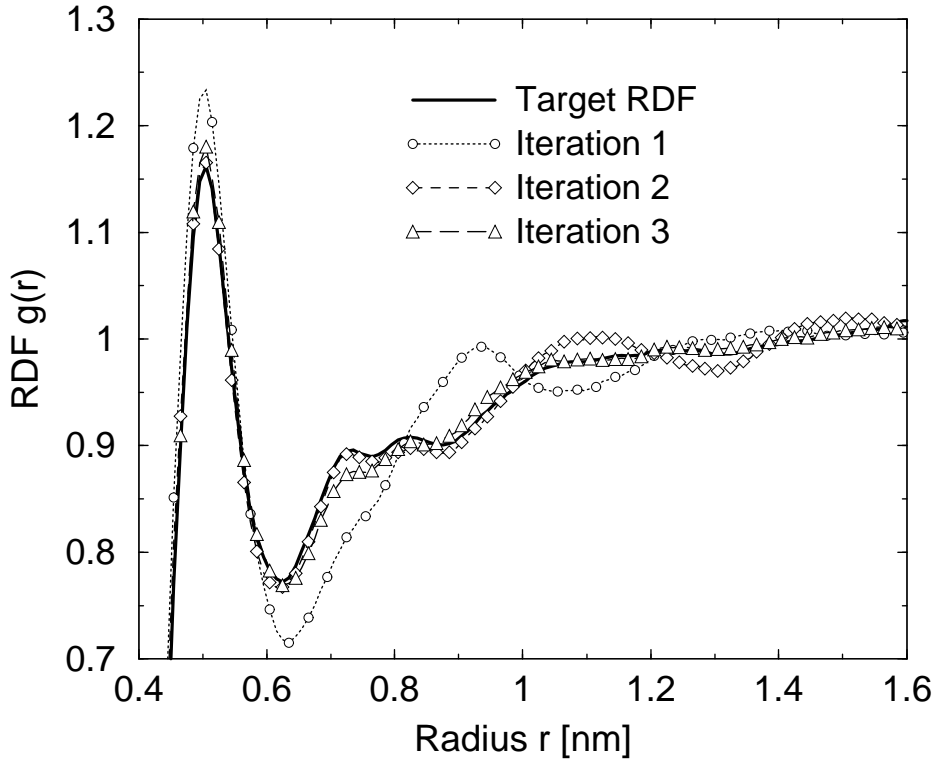}
  \end{center}
  \caption{Optimization of the intermolecular RDF for a melt of poly
  (isoprene) 9-mers by way of the iterative Boltzmann method. The RDF of
  iteration 1 corresponds to the initial potential guess, the direct
  Boltzmann inversion of the intermolecular target RDF. The quality of
  the trial RDFs improves very fast. After the third iteration (shown),
  the trial RDFs match the target within line thickness (not shown any more)}
  \label{pip-inter-it}
\end{figure}
The RDF converged quickly. The first three
iterations are shown in Figure~\ref{pip-inter-it}. The curve of
iteration $1$ is identical with the one shown in Figure~\ref{pip-prelim-opt}. 
It is far off the target curve at most distances.  Already the second
iteration $g_2(r)$, however, produces very good results up to a
distance of around $1$~nm. The third iteration $g_3(r)$ barely
deviates from the target curve. Convergence can be measured 
quantitatively by evaluating of the following merit function
($r_{\mbox{max}} = 1.5$~nm):
\begin{equation} \label{e:meritrdf}
f_{\text{target}} = \int w(r) \left(g(r) - g_j(r)\right)^2  dr \;.
\end{equation}
As a weighting function, we apply $w(r)=exp(-r/\sigma)$, in order to
penalize deviations at small distances more strongly. The results
are as follows:
\begin{verbatim}
#         f_target
Step 1    0.109621
Step 2    0.0107148
Step 3    0.0063149
Step 4    0.00223776
Step 5    0.00159995
Step 6    0.00144287
Step 7    0.00166992
Step 8    0.00128543
Step 9    0.00147876
\end{verbatim}
Looking at the merit function we can see that the RDF converged
after about five iterations. After that, the
deviations must be regarded as pure noise within our statistical
accuracy. One can also see that the potential solution remains stable.
A close look at Figure~\ref{pip-inter-it} also reveals that the curves converge
from short distances to large distances: areas which were mimicked
very accurately in an earlier iteration could be off in a
subsequent one, if there were areas at shorter distances which did
not match yet. This reflects the fact that changes in the
potential are non-local: short-range changes might have an impact at larger
distances, too. The final tabulated potential is shown in
Figure~\ref{pip-opt-results}.

\subsubsection{Simplex vs.\ Iterative Boltzmann Optimization}
Until recently, CG force field optimizations were also done 
by making use of the simplex algorithm in our
group~\cite{faller99,reith_crc02}. This procedure can be applied
when manipulating parameters of analytical force field models.
Typically, those contain several intra- and intermolecular terms,
each of which consists of a specific set of parameters, say $p_1,
\dots, p_n$.  For the iterative process, a single-valued merit
function is needed to evaluate the actual set of parameters and,
hence, the corresponding potential trial functions:
\begin{equation}
  \label{eq:opt_gen}
  f_{\text{target}} = f\left( \{ p_{k} \} \right) \longrightarrow \mbox{min!}
\end{equation}
The way to create new sets of parameters in order to minimize the
merit function can then be done by the simplex
algorithm~\cite{press92}. In case of the optimization of
structural properties, the merit function of
Equation~\ref{e:meritrdf} was utilized~\cite{reith00s}.  In
general, target values can come either from experimental data or
from simulations on a different length scale.  So far we used
thermodynamic state observables\cite{faller99} as well as continuous
functions as the radial distribution
function~\cite{meyer00,reith01phde}.\\

In order to compare the
capability of the simplex method compared to the iterative
Boltzmann method, we attempted an independent optimization with the
first one.
\begin{figure}[hbtp]
  \begin{center}
    \includegraphics[width=0.9\linewidth]{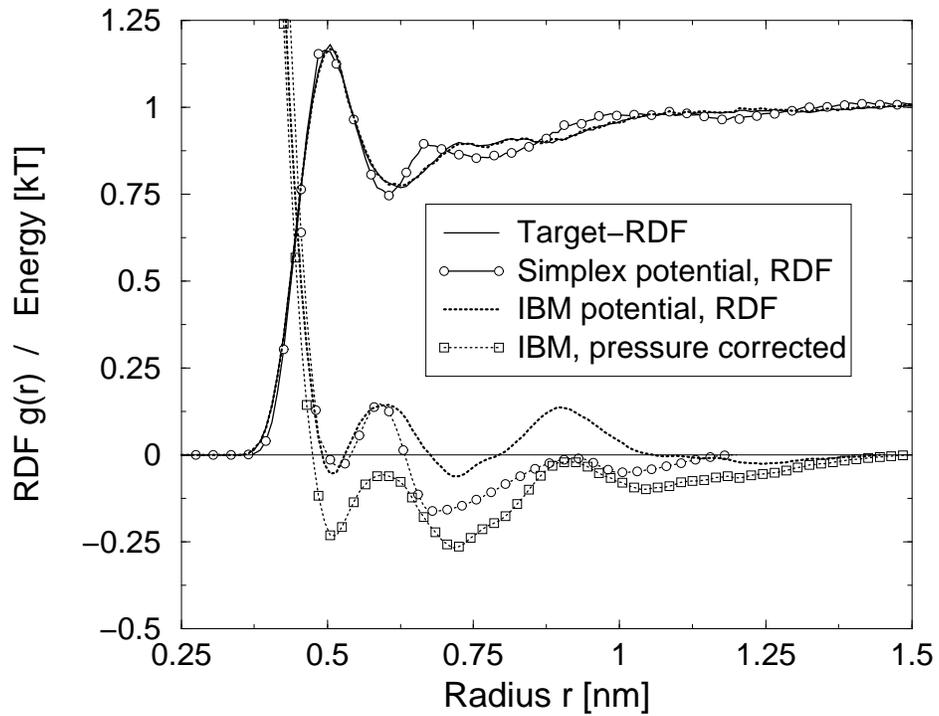}
  \end{center}
  \caption{Converged RDFs for mesoscopic simulations of a poly (isoprene)
    melt. Using the simplex algorithm, the area  $r \gtrsim 0.6$~nm could only
    qualitatively match the target RDF. In turn, the iterative
    Boltzmann-Inversion leads to a trial RDF which is identical to the target
    within line thickness. The last curve corresponds a the pressure-corrected
    result.}
  \label{pip-opt-results}
\end{figure}
We constructed the piecewise analytical trial potential for the
simplex optimization as follows (ignoring an additional shift to
$V=0$ at $r_{\rm cut}$):
\begin{equation}
  V_{\rm non-bonded}(r) = \left\{
    \begin{array}{ll}
      \varepsilon_1\,\left( \left({\sigma_1 \over r}\right)^8
        - \left({\sigma_1 \over r}\right)^6 \right) + \varepsilon_2
      &  r < \sigma_1 \\
      \varepsilon_2\,\left(\sin{(\sigma_1-r) \pi \over 2(\sigma_2-\sigma_1) }
      \right) + \varepsilon_2
      &  \sigma_1 \le r < \sigma_2 \\
      \varepsilon_3\,\left(\cos{(r-\sigma_2) \pi \over \sigma_3-\sigma_2 }
        - 1 \right)
      &  \sigma_2 \le r < \sigma_3 \\
      \varepsilon_4\,\left(-\cos{(r-\sigma_3) \pi \over \sigma_4-\sigma_3 }
        + 1 \right) - 2\varepsilon_3
      &  \sigma_3 \le r < \sigma_4\\
      \varepsilon_5\,\left(\cos{(r-\sigma_4) \pi \over \sigma_5-\sigma_4 }
        - 1 \right) - 2 (\varepsilon_3 - \varepsilon_4)
      &  \sigma_4 \le r < \sigma_5 \\
      \varepsilon_6\,\left(-\cos{(r-\sigma_5) \pi \over \sigma_6-\sigma_5 }
        + 1 \right) - 2 (\varepsilon_3 - \varepsilon_4 +
        \varepsilon_5)
      &  \sigma_5 \le r < r_{\rm cut} \; . \\
    \end{array} \right.
  \label{eq:pip-nb-simplex}
\end{equation}
Figure~\ref{pip-opt-results} proves that this formula, however 
complicated and lengthly, was
successfully adapted according to the needs imposed by the shape
of the target RDF: it comprises the possibility to generate a
third potential well. The idea to apply a third well (and its
position) came from the visual inspection of the
Boltzmann-inverted potential and its corresponding 
RDF (Figure~\ref{pip-prelim-opt}). The shape of the RDF up to
$r \approx 0.625$ nm indicates that the first potential well is a
bit too deep but still meaningful. In the subsequent area ($0.625$
nm $< r < 0.85$ nm), the trial RDF lies below the target RDF.
Therefore, we presumed a second potential well at around $0.7$ nm
would be needed to correct for this shortcoming. A similar
observation was made for the area $r > 0.85$ nm and we tried a
first guess well centered at $r = 1.1$ nm. These rough estimates
were utilized to initialize a simplex optimization. The optimum,
as shown in Figure~\ref{pip-opt-results}, fulfilled the condition
$f_{\text{target}} < 0.02$ and was reached after 150 iteration
steps. More details of the optimization procedure are given in
reference~\cite{reith01phde}. The final parameter set is listed in
Table~\ref{tab:PI_parms}. \\

In summary, the simplex algorithm has also been proven to generate acceptable
results. However, much experience, time and physical intuition were needed to
come to this final result.  Its quality is not as good as the quality of the
Boltzmann-inverted force field, although there are many similarities between
them, observing Figure~\ref{pip-opt-results}.  Therefore, its applicability
for structural problems seems to be inferior compared to the new inverted
Boltzmann scheme.  In retrospect it appears that the exponential weighting
function used in the merit function does not punish the long range deviation
in $g(r)$ enough. Thus, the difference in the merit function between the two
optimziation methods is small, although the inverted Boltzmann scheme  produces
visibly better results for second and third peaks in the correlation hole
regime.
%
\begin{table}[htbp]
\vspace*{3cm}
  \begin{center}
    \caption{Non-bonded interaction force field parameters for the
      coarse-grained model of a poly
      (isoprene) melt, according to Equation~\ref{eq:pip-nb-simplex},
      with $\sigma$ in nm and $\varepsilon$ in kT.}
    \begin{tabular}{cccccccccccc}
$\sigma_{1}$ & $\varepsilon_{1}$ & $\sigma_{2}$ & $\varepsilon_{2}$ %
& $\sigma_{3}$ & $\varepsilon_{3}$ & $\sigma_{4}$ & $\varepsilon_{4}$ %
& $\sigma_{5}$ & $\varepsilon_{5}$ & $\sigma_{6}=r_{\rm cut}$ & %
$\varepsilon_{6}$\\
& & & & & & & & & & &\\[-6pt] \hline\hline
& & & & & & & & & & &\\[-6pt]
& & & & & & & & & & &\\
0.485 & 1.1 & 0.61 & 0.09 & 0.724 & 0.228 %
& 0.927 & 0.24 & 1.08 & 0.038 & 1.3 & 0.05 \\[5pt]
    \end{tabular}
    \vspace*{0.3cm}
    \label{tab:PI_parms}
  \end{center}
\end{table}
%

%
\subsubsection{Pressure correction}
So far, we were just concerned with the structural optimization of
the PI melt system. Our best potential 
(Figure~\ref{pip-opt-results}), however, has a positive
pressure of $p^* = 1.92$ (given in reduced units~\cite{All87}). 
This does not reflect the ambient conditions of the parent atomistic 
system~\cite{faller01a}. 
This is a consequence of the simulation being run at constant
volume and thermodynamic properties not being present in the merit
function. As a proof of concept, that pressure correction is possible,
we tried to post-optimize the mesoscopic system without
lowering the quality of radial distribution function. The potential
of a neutral system is always attractive at long range due to the
dispersion interactions. Consequently, we chose an attractive linear 
tail function, as a weak perturbation to the potential previously
optimized without pressure correction:
\begin{equation}
  \Delta V_{\rm lin}(r) \; = \; 
  A \left( 1 - \frac{r}{r_{\rm cutoff}} \right) ,
\end{equation}
with $A = -0.1k_BT$. 
The correction fulfills the following essential conditions: 
$\Delta V_{\rm lin}(0) = A$ and
$\Delta V_{\rm lin}(r_{\rm cutoff}) = 0$. The corrected potential 
was then taken as initial guess of a re-optimization of the potential
against the structure using the iterative Boltzmann method. After
that, the pressure was re-evaluated and the procedure continued
until $p^* < 0.005$. This condition was reached after 10
iteration cycles. The pressure-corrected potential is also shown
in Figure~\ref{pip-opt-results}. As suspected, the shape and
with it, the forces, are very similar to the force
field optimized without pressure corrections. The pressure correction 
manifests itself mostly in the potential region beyond $r > 1$ nm, which
is not very important for the structural fit. It shifts the whole
potentials downward, thus providing for the previously missing
long-range attraction.
\subsection{Poly (isoprene) in cyclohexane: Solution optimization and comparison
to the melt}
Our atomistic PI systems were highly diluted and only system S3
contained more than one PI 15-mer. Since statistical accuracy was not
sufficient to analyze intermolecular radial distribution functions, we 
chose to optimize intramolucular degrees of freedom only. However, 
1-4 interactions and higher were considered in the same way as 
intermolecular contacts and a purely repulsive potential was applied 
for all those interactions to complete the force field 
properly. The repulsive potential was based on information of
the melt simulations: The repulsive part between $r=0$ and the mimimum 
was cut off at the minimum and 
shifted to zero. The optimization of the intramolecular part was done 
in the same order as described before (cf.\ section~\ref{melt-opt})
So, the stretching potential $V_{\rm stretch}$ was again our starting
point. As expected for this rigid degree of freedom, the distribution
of the bond distances was close to a Gaussian shape and it could be 
mimicked with almost identical parameters as in the melt system. 
This means, it is almost independent of the concentration
of the system, in contrast to all other degrees of freedom. For
the bending potential $V_{\rm bend}$, the Boltzmann-inverted
distribution $P(\alpha)$ was taken as initial guess, according to
Equation~\ref{v0}. After five iterations, the agreement
shown in Figure~\ref{pip-intra-distr} was achieved. The
coincidence of the main peak around 160$^o$ was still not perfect.
This originated from the metric-tensor corrections (division by
$\sin \alpha$), which magnify potential corrections too
strongly. However, the result is still satisfactory. For the
torsions, the curves were again acceptable without any
optimization. This shows the value of meaningful initial
guesses and the accuracy of Boltzmann-inverted potentials, at
least for some degrees of freedom. Taking the intramolecular
potentials together as force field, the resulting RDF matches the
target RDF very well. Meaningful static properties could be
derived, as published elsewhere~\cite{fallerreith02}.

\begin{figure}[tbp]
  \begin{center}
    \includegraphics[width=0.7\linewidth]{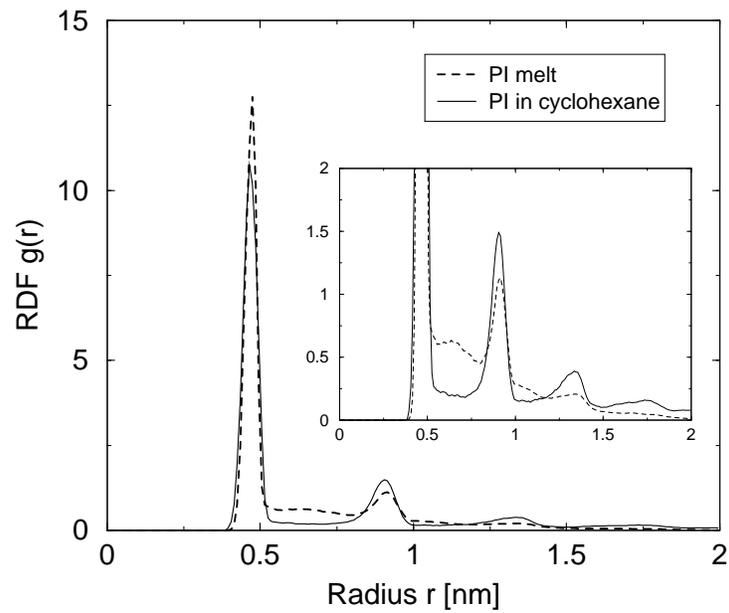}
  \end{center}
  \caption{Comparison of intrachain radial distribution functions (RDFs) for
    super-atoms as obtained from mesoscopic simulations of a poly (isoprene)
    melt and dilute solution. The integrals under both curves have been normalized 
    to $1$ (with integral cutoff at $r = 2$~nm.) in order to compare relative peak
    heights.}
  \label{rdf-comp}
\end{figure}
We are now also in the position to compare the intrachain RDFs of the
mesoscopic simulations of a PI melt and a solution system.
Figure~\ref{rdf-comp} shows both curves with integrals normalized to $1$
($r_{\text cut} = 2$~nm.) in order to compare relative peak heights. The fact
that the intramolecular distribution functions are different is also reflected
here: the peak positions are the same, but the relative intensity varies. In
solution, the atomistic \textit{trans} configuration is mesoscopically
reflected by a high population of stretched-out states, due to good solvent
conditions.

%
%
\section{Discussion and Conclusions}
%
%
In this contribution, we introduced an iterative method for potential
inversion from distribution functions for polymer systems. However, this
scheme could also be utilized for other chemical systems in which
intramolecular connectivity has to be taken into account, such as
low-molecular solvents or proteins. Basic tests on two monoatomic liquids 
(Lennard-Jones and WCA) with known solutions showed, that (1) in all cases the
algorithm produced numerical potentials which led to radial distribution 
functions undistinguishable from the target, (2) the algorithm was robust and
deviations between the iterative solution and the exact potential were minor,
and (3) that the \textit{a priori} unknown range of the potential is important.
Note also, that taking numerical inaccuracies into account, the forces 
generated by different useful potentials are very similar as observed in the 
corresponding graphs. As a conclusion, we would advocate to start with a range 
as small as possible and increase it stepwise until deviations of the RDFs from 
their targets are satisfactory at all ranges. 

The algorithm was applied to trans-1,4 poly (isoprene) oligomers in two different
situations: melt and solution. Although the conformational differences between
them inherited from the parent atomistic reference calculations still need to be
understood more deeply, it is clear that iterative Boltzmann inversion can 
successfully coarse-grain both of them. The CG force fields turn out to be similar, 
but the clearly distinct from each other. For the melt system, it was also shown 
that the pressure could be post-optimized by adding a weak attractive perturbation 
potential without disturbing the short-range structure noticably. This example
illustrates nicely that, due to the neglect of (different) degrees of freedom
during the CG procedure, the resulting CG potentials necessarily depend on the 
state of the polymer, here its environment: Even the intramolecular parts of the
potential differ in melt and solution.

\section*{Acknowledgements}
We are indebted to Roland Faller for providing the atomistic force
fields and part of the atomistic data of poly (isoprene). Hendrik
Meyer is deeply acknowledged for sharing his valuable ideas with
us. We also want to thank Cameron Abrams for fruitful discussions.

%
%

\clearpage


%
%
%
%
%
%
\clearpage

%
%

%
%
%
%
%
%
%
%

\end{document}